\UseRawInputEncoding

\documentclass{ws-ijmpd}
\usepackage[super,compress]{cite}
\begin{document}

\markboth{Zi-Hua Weng}
{Frequencies of astrophysical jets and gravitational strengths in the octonion spaces}

%
\catchline{}{}{}{}{}
%

\title{Frequencies of astrophysical jets and gravitational strengths in the octonion spaces
}

\author{Zi-Hua Weng
}

\address{
School of Aerospace Engineering, Xiamen University, China \\
College of Physical Science and Technology, Xiamen University, China
\\
xmuwzh@xmu.edu.cn}



\maketitle

\begin{history}
\received{Day Month Year}
\revised{Day Month Year}
\end{history}

\begin{abstract}
  The paper focuses on applying the octonions to explore the electromagnetic and gravitational equations in the presence of some material media, exploring the frequencies of astrophysical jets. Maxwell was the first to use the algebra of quaternions to describe the electromagnetic equations. This method encourages scholars in adopting the quaternions and octonions to study the physical properties of electromagnetic and gravitational fields, including the field strength, field source, linear momentum, and angular momentum and so forth. In the paper, the field strength and angular momentum in the vacuum can be combined together to become one new physical quantity, that is, the composite field strength within the material media. Substituting the latter for the field strength in the vacuum will deduce the field equations within material media, including the electromagnetic and gravitational equations in the presence of some material media. In terms of the electromagnetic fields, the electromagnetic equations in the presence of some electromagnetic media are able to explore a few new physical properties of electromagnetic media. Especially, in case the magnetic flux density and magnetization intensity both fluctuate at a single-frequency, their frequencies must be identical to each other within electromagnetic media. In some extreme cases, the electromagnetic equations within electromagnetic media will be degenerated into the Maxwell's equations in the vacuum. The above reveals that the electromagnetic equations within electromagnetic media are capable of extending the scope of application of electromagnetic theory. For the gravitational fields, there are some similar inferences and conclusions within gravitational media. Further they can be utilized to research the frequencies of astrophysical jets.
\end{abstract}


\ccode{PACS numbers: 07.55.Jg; 03.50.De; 02.10.De; 04.50.-h; 11.10.Kk.}


\section{\label{sec:level1}Introduction}

Does each of field equations need to be extended, when the electromagnetic equations are extended from the vacuum into the electromagnetic media? Is it necessary to choose one new gauge condition, rather than the gauge condition in the vacuum, for the electromagnetic equations in the presence of the electromagnetic media? How do we deal with some similar cases in the gravitational fields with gravitational media? For a long time, these puzzles have attracted and tempted scholars' attentions, perplexing some scholars all the time. Until the emergence of the electromagnetic and gravitational field theory described with the octonions (octonion field theory, for short temporarily), the above difficult problems are solved to some extent. In the octonion field theory, when the electromagnetic equations are extended from the vacuum into the electromagnetic media, each of field equations will be extended, and it is necessary to choose one new gauge condition. Similar treatments should be made for the gravitational fields. As a result, the octonion field theory is able to explore a few new physical properties relevant to the electromagnetic and gravitational media.

After analyzing and summarizing the relevant literatures over one hundred years before his time, Maxwell \cite{maxwell} published his works related with the electromagnetic fields in 1876. In his works, he introduced the quaternions and vector terminology to describe the physical properties of electromagnetic fields. Making use of twenty scalar equations, Maxwell concluded the basic equations of the classical electrodynamics, including the electromagnetic equations, electromagnetic force, and current continuity equation. In particular, Maxwell introduced the concept of displacement current, making the electromagnetic equations to be complete and self-consistent. In 1887 Hertz \cite{hertz} validated the existence of displacement current in experiments.

The algebra of quaternions, used in the Maxwell's works, was invented by Hamilton \cite{hamilton} in 1843. Cayley \cite{cayley} and Graves invented respectively the algebra of octonions in 1844. Obviously the quaternion is one great invention. However the engineers, in the nineteenth century, didn't think the quaternions were convenient enough to use. The quaternions were gradually decomposed into the vectors and scalars. Subsequently Maxwell applied simultaneously the quaternions and vector terminology to explore the basic equations of the classical electrodynamics. In 1893, Heaviside substituted drastically the vectors and scalars for quaternions, reformulating successfully the basic equations of the classical electrodynamics, reducing effectively the difficulty of electromagnetic theory and even gravitational theory \cite{heaviside1,heaviside2}. Since then, the electromagnetic theory has been widely used by scholars and others.

As time goes on and the experiment evolves, the electromagnetic equations are also extended from the vacuum to the electromagnetic medium. For this reason, the scholars \cite{kansu1} introduced the equation of electromagnetic properties of medium, or the equation of state of the electromagnetic medium. Apparently, it enlarges the scope of application of the electromagnetic equations.

In the twentieth century, it had been found that the method of vector terminology occupies a few fatal weaknesses. For instance, sometimes there will be the failure condition of `all elements of a matrix are zero simultaneously', in some pivotal matrices relevant to the satellite orbit control, by means of the vector terminology. However, no such failure condition would occur, in case these pivotal matrices are described by quaternions. It means that the method of vector terminology is unable to completely replace that of quaternions. Compared with the former, the latter possesses several irreplaceable advantages. This unexpected discovery and other investigations prompted scholars to return and continue to apply the algebra of quaternions to study electromagnetic theory. In the method of quaternions, the electromagnetic equations can be described by one single-equation \cite{honig, singh}. In other words, the four equations of the Maxwell's equations are able to be rewritten as one single quaternionic equation.

This method encourages scholars \cite{rajput} to apply the quaternions and octonions to study the physical properties of electromagnetic and gravitational fields. Negi et al. \cite{negi} deduced the quaternionic formalism for electromagnetic equations. Majernik \cite{majernik} presented the quaternionic formulation of classical Maxwell-like field equations using quaternions. Grusky et al. \cite{grusky} reduced the Maxwell's equations for the time-dependent electromagnetic field in a homogeneous chiral medium into a single quaternionic equation. Morita \cite{morita} applied the quaternions to describe the Lorentz group and Dirac theory. Demir et al. \cite{demir1,demir2} proposed an alternative method to formulate the classical and generalized electromagnetism, in the case of the existence of magnetic monopoles \cite{demir3} and massive photons. Rawat et al. \cite{rawat} discussed the gravitational and electromagnetic fields with the quaternions. Kansu et al. \cite{kansu2} proposed the generalized description of electromagnetism and linear gravity, based on the combined dual numbers and complex quaternion algebra. Moffat \cite{moffat} showed an eight-dimensional Riemannian geometry to be the basis of a nonsymmetric theory of gravitation. De Leo et al. \cite{deleo} proposed a coupled eigenvalue problem for octonionic operators, discussing the hermiticity of such operators. Anastassiu et al. \cite{anastassiu} applied the quaternions into the derivation of analytical solutions of Maxwell's equations. Gogberashvili \cite{gogberashvili} derived the Dirac's operator and Maxwell's equations in vacuum from the algebra of split octonions. Mironov et al. \cite{mironov} described the electromagnetic field in a vacuum by means of the split octonions. Chanyal et al. \cite{chanyal} made an attempt to reformulate the generalized field equation of dyons in terms of octonion variables. Furui \cite{furui} calculated the axial current and two vector currents triangle diagram of Bardeen in terms of the octonions.

The algebra of octonions was introduced by Graves and Cayley independently. This type of octonions can be called as the classical octonions. What the paper discusses is the classical octonions, rather than the non-classical octonions, including hyperbolic-octonions, split-octonions, pseudo-octonions, Cartan¡¯s octonions, and others.

In the paper, the algebra of octonions is capable of describing simultaneously the physical properties of electromagnetic and gravitational fields. The octonion space $\mathbb{O}$ can be separated into a few subspaces independent of each other, including $\mathbb{H}_g$ and $\mathbb{H}_{em}$ . The subspace $\mathbb{H}_g$ can be applied to depict the gravitational fields, while the second subspace $\mathbb{H}_{em}$ may be utilized to describe the electromagnetic fields. It means that the application of octonion space $\mathbb{O}$ is able to research simultaneously the electromagnetic equations, gravitational equations, and angular momentum and so forth within the vacuum or material media.

Compared with the classical electromagnetic theory, we may find some characteristics of the octonion field theory as follows.

1) Field strength. In the electromagnetic equations for the vacuum, the electromagnetic strength consists merely of the electric field intensity $\textbf{E}$ and magnetic induction intensity $\textbf{B}$ . In the electromagnetic equations for the electromagnetic media, can the electromagnetic strength include simply the electric induction intensity $\textbf{D}$ and magnetic field intensity $\textbf{H}$ ? In the classical electromagnetic theory, the electromagnetic equations for the electromagnetic media have to contain $\textbf{B}$ , besides $\textbf{D}$ and $\textbf{H}$ . Contrastively, in the octonion field theory, the electromagnetic equations for the electromagnetic media are able to consist merely of $\textbf{D}$ and $\textbf{H}$ , simplifying the electromagnetic equations in the electromagnetic medium.

2) Gauss equation. In terms of the classical electromagnetic theory, there is one Gauss's law for magnetism, in the electromagnetic equations for the vacuum. Also there is another Gauss's law for magnetism, in the electromagnetic equations for the electromagnetic medium. Why do the two Gauss's laws for magnetism have to be identical completely? Some scholars doubt the rationality of Gauss's law for magnetism, in the electromagnetic equations for the electromagnetic medium. However, in the octonion field theory, the Gauss's law for magnetism, in the electromagnetic equations for the electromagnetic medium, is allowed to be different from that in the electromagnetic equations for the vacuum, expanding the applicability of electromagnetic equations.

3) Gauge condition. In the electromagnetic equations for the vacuum, the steady magnetic fields satisfy the Coulomb gauge condition, while the transient electromagnetic fields obey the Lorentz gauge condition. In terms of the classical electromagnetic theory, in the electromagnetic equations for the electromagnetic medium, why do the transient electromagnetic fields still obey the Lorentz gauge condition, rather than one new gauge condition? It reveals that the electromagnetic equations may not be extended completely from the vacuum into the electromagnetic medium, in the classical electromagnetic theory. And there are some aspects to be improved in this theory. Nevertheless, in terms of the octonion field theory, the gauge condition for the electromagnetic equations in the electromagnetic medium is different from that in the vacuum, simplifying the electromagnetic equations.

Similarly, compared with the classical gravitational theory, one can find some characteristics relevant to the gravitational fields in the octonion field theory.

In the paper, the octonion field strength and angular momentum in the vacuum can be combined together to become one new physical quantity. Substituting the latter for the octonion field strength in the vacuum will deduce the electromagnetic equations and gravitational equations in the material media. The electromagnetic equations for the electromagnetic medium, in the octonion field theory, can be degenerated into that in the classical electromagnetic theory. Similarly, the gravitational equations for the gravitational medium, in the octonion field theory, can be degenerated into that in the classical gravitational theory.

\begin{table}[h]
\tbl{Some physical quantities relevant to the electromagnetic and gravitational fields in two subspaces of octonion spaces.}
{\begin{tabular}{@{}ll@{}}
\hline\hline
subspace                                                  &  physical quantity                                                                                   \\
\hline
quaternion space, $\mathbb{H}_g (\textbf{i}_j)$           &  $\mathbb{R}_g$ , $\mathbb{V}_g$ , $\mathbb{Y}_g$ , $\mathbb{A}_g$ , $\mathbb{X}_g$ ,
                                                             $\mathbb{F}_g$ ,  $\mathbb{S}_g$ , $\mathbb{P}_g$ , $\mathbb{L}_g$ .                                \\
second subspace, $\mathbb{H}_{em} (\textbf{I}_j)$         &  $\mathbb{R}_e$ , $\mathbb{V}_e$ , $\mathbb{Y}_e$ , $\mathbb{A}_e$ , $\mathbb{X}_e$ ,
                                                             $\mathbb{F}_e$ , $\mathbb{S}_e$ , $\mathbb{P}_e$ , $\mathbb{L}_e$ .                                 \\
\hline\hline
\end{tabular} \label{tab:1}}
\end{table}

\section{Octonion space}

The octonion space $\mathbb{O}$ can be separated into a few subspaces independent of each other, including $\mathbb{H}_g$ and $\mathbb{H}_{em}$ . The subspace $\mathbb{H}_g$ is one quaternion space, which can be applied to depict the physical properties of gravitational fields. Meanwhile, the second subspace $\mathbb{H}_{em}$ may be utilized to describe the physical properties of electromagnetic fields (Table 1).

In the quaternion space $\mathbb{H}_g$ , the basis vector is $\textbf{i}_j$ , the coordinate is $r_j$ . The quaternion radius vector is, $\mathbb{R}_g = i r_0 \textbf{i}_0 + \Sigma r_k \textbf{i}_k$, and the quaternion velocity is, $\mathbb{V}_g = i v_0 \textbf{i}_0 + \Sigma v_k \textbf{i}_k$ .
Meanwhile, in the second subspace $\mathbb{H}_{em}$ , the basis vector is $\textbf{I}_j$, the coordinate is $R_j$ . The radius vector is, $\mathbb{R}_e = i R_0 \textbf{I}_0 + \Sigma R_k \textbf{I}_k$ , and the velocity is, $\mathbb{V}_e = i V_0 \textbf{I}_0 + \Sigma V_k \textbf{I}_k$. The octonion radius vector is, $\mathbb{R} = \mathbb{R}_g + k_{eg} \mathbb{R}_e$ , and the octonion velocity is, $\mathbb{V} = \mathbb{V}_g + k_{eg} \mathbb{V}_e$ . Herein $\textbf{I}_j = \textbf{i}_j \circ \textbf{I}_0$ . $r_0 = v_0 t$ . $t$ is the time. $v_0$ is the speed of light. $\circ$ denotes the multiplication of octonions. $i$ is the imaginary unit. $k_{eg}$ is a coefficient, meeting the demand for the dimensional homogeneity in the physics. $r_j$ , $v_j$, $R_j$ , and $V_j$ are all real. $\textbf{i}_0 = 1$. $\textbf{i}_k^2 = - 1$. $\textbf{I}_j^2 = - 1$. $j = 0, 1, 2, 3$. $k =1, 2, 3$.

In the field theories, the quaternion space $\mathbb{H}_g$ is fit for describing the physical properties of gravitational fields, while the second subspace $\mathbb{H}_{em}$ is propitious to depict that of electromagnetic fields. It implies that the octonion space $\mathbb{O}$ is able to bewrite simultaneously the physical properties of gravitational and electromagnetic fields. In the quaternion space $\mathbb{H}_g$ , the gravitational potential is, $\mathbb{A}_g = i a_0 \textbf{i}_0 + \Sigma a_k \textbf{i}_k$, the gravitational strength is, $\mathbb{F}_g = f_0 \textbf{i}_0 + \Sigma f_k \textbf{i}_k$ , and the gravitational source is, $\mathbb{S}_g = i s_0 \textbf{i}_0 + \Sigma s_k \textbf{i}_k$ . Meanwhile, in the second subspace $\mathbb{H}_{em}$ , the electromagnetic potential is, $\mathbb{A}_e = i A_0 \textbf{I}_0 + \Sigma A_k \textbf{I}_k$ , the electromagnetic strength is, $\mathbb{F}_e = F_0 \textbf{I}_0 + \Sigma F_k \textbf{I}_k$, and the electromagnetic source is, $\mathbb{S}_e = i S_0 \textbf{I}_0 + \Sigma S_k \textbf{I}_k$ . Herein $\mathbb{F}_g = f_0 + i \textbf{g} / v_0 + \textbf{b}$. $\mathbb{F}_e = \textbf{F}_0 + i \textbf{E} / v_0 + \textbf{B}$ . $\textbf{g}$ is the gravitational acceleration, and $\textbf{b}$ is called as the gravitational precessional-angular-velocity temporarily \cite{weng1}. $\textbf{F}_0 = F_0 \textbf{I}_0$ . $\textbf{S}_0 = S_0 \textbf{I}_0$ . $\textbf{S} = \Sigma S_k \textbf{I}_k$. $\textbf{s} = \Sigma s_k \textbf{i}_k$ . $a_j$, $A_j$ , $f_0$, $F_0$, and $S_j$ are all real. $f_k$ and $F_k$ both are complex numbers.

In the octonion space $\mathbb{O}$ , the gravitational potential and electromagnetic potential are able to combine together to become the octonion field potential, $\mathbb{A} = \mathbb{A}_g + k_{eg} \mathbb{A}_e$ . The gravitational strength and electromagnetic strength can combine together to become the octonion field strength, $\mathbb{F} = \mathbb{F}_g + k_{eg} \mathbb{F}_e$ . From the octonion field strength $\mathbb{F}$ , one can define the octonion field source as follows,
\begin{eqnarray}
\mu \mathbb{S} = - ( i \mathbb{F} / v_0 + \lozenge )^\ast \circ \mathbb{F}  ~,
\end{eqnarray}
where $\mathbb{F} = \lozenge \circ \mathbb{A}$ . $\mu$ is a coefficient. $\ast$ stands for the octonion conjugate. $\lozenge$ is the quaternion operator. $\lozenge = \partial_0 + \nabla $ . $\nabla = \Sigma \textbf{i}_k \partial_k$ . $\partial_j = \partial / \partial r_j$ .

In the above, the gravitational source $\mathbb{S}_g$ and electromagnetic source $\mathbb{S}_e$ can be combined together to become the octonion field source $\mathbb{S}$ ,
\begin{eqnarray}
\mu \mathbb{S} = \mu_g \mathbb{S}_g + k_{eg} \mu_e \mathbb{S}_e -  i \mathbb{F}^\ast \circ \mathbb{F} / v_0 ~ ,
\end{eqnarray}
where $\mu_g$ is the gravitational field constant, while $\mu_e$ is electromagnetic field constant. $k_{eg}^2 = \mu_g / \mu_e$ .

From the octonion field source $\mathbb{S}$ , we can define the octonion linear momentum, $\mathbb{P} = \mu \mathbb{S} / \mu_g$ . And it can be separated into two parts,
\begin{eqnarray}
\mathbb{P} = \mathbb{P}_g + k_{eg} \mathbb{P}_e  ~ ,
\end{eqnarray}
where $\mathbb{P}_g = \mathbb{S}_g - i \mathbb{F}^\ast \circ \mathbb{F} / ( v_0 \mu_g )$ , $\mathbb{P}_e = \mu_e \mathbb{S}_e / \mu_g$ . $\mathbb{P}_g$ is the component of $\mathbb{P}$ in the quaternion space $\mathbb{H}_g$ , while $\mathbb{P}_e$ is the component of $\mathbb{P}$ in the second subspace $\mathbb{H}_{em}$ .

It is able to define the octonion field potential, $\mathbb{A} = i \lozenge^\times \circ \mathbb{X}$ , from the octonion integrating function of field potential $\mathbb{X}$ . $\times$ denotes the complex conjugate. And the latter can be decomposed into two parts,
\begin{eqnarray}
\mathbb{X} = \mathbb{X}_g + k_{eg} \mathbb{X}_e  ~ ,
\end{eqnarray}
where $\mathbb{X}_g$ is the component of $\mathbb{X}$ in the quaternion space $\mathbb{H}_g$ , while $\mathbb{X}_e$ is the component of $\mathbb{X}$ in the second subspace $\mathbb{H}_{em}$ .

Further, from the octonion linear momentum $\mathbb{P}$ , one can define the octonion angular momentum as follows,
\begin{eqnarray}
\mathbb{L} = ( \mathbb{R} + k_{rx} \mathbb{X} )^\times \circ \mathbb{P}  ~ ,
\end{eqnarray}
where $k_{rx} = 1 / v_0$ is a coefficient, to meet the demand for the dimensional homogeneity. Further one may define the octonion torque $\mathbb{W}$ and force $\mathbb{N}$ in the octonion space (Table \ref{tab:2}).

\begin{table}[h]
\tbl{Some equations of the gravitational and electromagnetic fields in the vacuum.}
{\begin{tabular}{@{}ll@{}}
\hline\hline
physical quantity             &   definition                                                                             \\
\hline
field potential               &   $\mathbb{A} = i \lozenge^\times \circ \mathbb{X}$                                      \\
field strength                &   $\mathbb{F} = \lozenge \circ \mathbb{A}$                                               \\
field source                  &   $\mu \mathbb{S} = - ( i \mathbb{F} / v_0 + \lozenge )^\ast \circ \mathbb{F}$           \\
linear momentum               &   $\mathbb{P} = \mu \mathbb{S} / \mu_g$                                                  \\
angular momentum              &   $\mathbb{L} = ( \mathbb{R} + k_{rx} \mathbb{X} )^\times \circ \mathbb{P} $             \\
octonion torque               &   $\mathbb{W} = - v_0 ( i \mathbb{F} / v_0 + \lozenge ) \circ \mathbb{L}$                \\
octonion force                &   $\mathbb{N} = - ( i \mathbb{F} / v_0 + \lozenge ) \circ \mathbb{W}$                    \\
\hline\hline
\end{tabular} \label{tab:2}}
\end{table}

The octonion angular momentum $\mathbb{L}$ can be separated into two parts,
\begin{eqnarray}
\mathbb{L} = \mathbb{L}_g + k_{eg} \mathbb{L}_e  ~ ,
\end{eqnarray}
where $\mathbb{L}_g$ is the component of $\mathbb{L}$ in the quaternion space $\mathbb{H}_g$ , while $\mathbb{L}_e$ is the component of $\mathbb{L}$ in the second subspace $\mathbb{H}_{em}$ . $\mathbb{L}_g = L_{10} + i \textbf{L}_1^i + \textbf{L}_1$ , $\mathbb{L}_e = \textbf{L}_{20} + i \textbf{L}_2^i + \textbf{L}_2$ . $\textbf{L}_1$ includes the angular momentum density. $\textbf{L}_2^i$ is in direct proportion to the electric moment density, and $\textbf{L}_2$ covers the magnetic moment density. $\textbf{L}_1^i = \Sigma L_{1k}^i \textbf{i}_k$. $\textbf{L}_2^i = \Sigma L_{2k}^i \textbf{I}_k$. $\textbf{L}_1 = \Sigma L_{1k} \textbf{i}_k$. $\textbf{L}_2 = \Sigma L_{2k} \textbf{I}_k$ . $\textbf{L}_{20} = L_{20} \textbf{I}_0$ . $L_{1k}^i$ , $L_{2k}^i$ , $L_{1j}$ , and $L_{2j}$ are all real.

The preceding context deals with the investigation of the gravitational and electromagnetic fields in the vacuum \cite{weng2}. As we all know, a few physical properties of the gravitational and electromagnetic fields in the material media may be different from that in the vacuum. Apparently, the method for exploring the gravitational and electromagnetic fields in the vacuum can be expanded into that in the material media.

\section{Field equations}

In the Maxwell's electromagnetic theory, the compositions of the electric induction intensity $\textbf{D}$ and magnetic field intensity $\textbf{H}$ give an important enlightenment to this paper. The electric field intensity $\textbf{E}$ and electric polarization intensity compose the electric induction intensity $\textbf{D}$ in the electric media. And the magnetic induction intensity $\textbf{B}$ and magnetization intensity constitute the magnetic field intensity $\textbf{H}$ in the magnetic media. The introduction of the electric induction intensity $\textbf{D}$ and magnetic field intensity $\textbf{H}$ enables us to explore the physical properties within the electromagnetic media.

This characteristics reveals that the octonion field strength $\mathbb{F}$ and octonion angular momentum $\mathbb{L}$ can be combined together to become one new physical quantity, $\mathbb{F}^+ = \mathbb{F} + k_{fl} \mathbb{L}$ , which is called as the composite field strength temporarily. In the octonion field theory, the octonion angular momentum $\mathbb{L}$ consists of the electric polarization intensity and magnetization intensity and others. By analogy with the definition of octonion field equations in the vacuum, it is able to define the composite field source, $\mu \mathbb{S}^+$ , from the quaternion operator $\lozenge$ and composite field strength,
\begin{eqnarray}
\mu \mathbb{S}^+ = - ( i \mathbb{F}^+ / v_0 + \lozenge )^\ast \circ \mathbb{F}^+  ~,
\label{equ:7}
\end{eqnarray}
where $\mu \mathbb{S}^+ = \mu \mathbb{S} + k_{fl} \mathbb{Z}$ . $\mathbb{S}^+$ and $\mathbb{Z}$ both are octonions. $\mathbb{Z}$ is one component of field sources that can merely be produced or existed in a material media. In case $k_{fl} \mathbb{L}$ and $k_{fl} \mathbb{Z}$ both can be neglected, the above will be degenerated into the definition of octonion field equations in the vacuum. $k_{fl}$ is a coefficient, to meet the demand for the dimensional homogeneity.

Through comparison and analysis, it is found that the component of field source, $\mathbb{Z}$ , is contrasted with the octonion torque $\mathbb{W}$ , for their distinct definitions. According to the definition, the component of field source, $\mathbb{Z}$ , includes the particles that can exist merely in material media. The above can be utilized to research the characteristics of electromagnetic and gravitational media.

The composite field strength $\mathbb{F}^+$ can be rewritten as,
\begin{eqnarray}
\mathbb{F} + k_{fl} \mathbb{L} = ( \mathbb{F}_g + k_{fl} \mathbb{L}_g ) + k_{eg} ( \mathbb{F}_e + k_{fl} \mathbb{L}_e )  ~ ,
\end{eqnarray}
or
\begin{eqnarray}
\mathbb{F}^+ = \mathbb{F}^+_g + k_{eg} \mathbb{F}^+_e  ~ ,
\end{eqnarray}
where $\mathbb{F}^+_g = \mathbb{F}_g + k_{fl} \mathbb{L}_g$ , $\mathbb{F}^+_e = \mathbb{F}_e + k_{fl} \mathbb{L}_e$ .

Meanwhile the composite field source, Eq.(\ref{equ:7}), will be written as,
\begin{eqnarray}
\mu \mathbb{S} + k_{fl} \mathbb{Z}
= ( \mu_g \mathbb{S}_g + k_{fl} \mathbb{Z}_g ) + k_{eg} ( \mu_e \mathbb{S}_e + k_{fl} \mathbb{Z}_e )
   - i ( \mathbb{F}^+ )^\ast \circ \mathbb{F}^+ / v_0  ~ ,
\end{eqnarray}
where $\mathbb{Z} = \mathbb{Z}_g + k_{eg} \mathbb{Z}_e$ . $\mathbb{Z}_g$ and $\mathbb{Z}_e$ are respectively the components of $\mathbb{Z}$ in the two subspaces, $\mathbb{H}_g$ and $\mathbb{H}_{em}$. Further one can define the composite linear momentum $\mathbb{P}^+$, angular momentum $\mathbb{L}^+$ , torque $\mathbb{W}^+$ , and force $\mathbb{N}^+$ in the octonion space (Table \ref{tab:3}).

According to the coefficient $k_{eg}$ , one can decompose the definition of composite field source into two parts,
\begin{eqnarray}
\mu_g \mathbb{S}^+_g = - \lozenge^\ast \circ \mathbb{F}^+_g     ~ ,
\label{equ:11}
\\
\mu_e \mathbb{S}^+_e = - \lozenge^\ast \circ \mathbb{F}^+_e     ~ ,
\label{equ:12}
\end{eqnarray}
where $\mu_g \mathbb{S}^+_g = \mu_g \mathbb{S}_g + k_{fl} \mathbb{Z}_g$ . $\mu_e \mathbb{S}^+_e = \mu_e \mathbb{S}_e + k_{fl} \mathbb{Z}_e$ . The former is the definition of gravitational equations for gravitational media in the quaternion space $\mathbb{H}_g$ , while the latter is that of electromagnetic equations for electromagnetic media in the second subspace $\mathbb{H}_{em}$ .

\begin{table}[h]
\tbl{Some equations of the gravitational and electromagnetic fields in the material media.}
{\begin{tabular}{@{}ll@{}}
\hline\hline
physical quantity                               &   definition                                                                                 \\
\hline
composite field strength                        &   $\mathbb{F}^+ = \mathbb{F} + k_{fl} \mathbb{L}$                                            \\
composite field source                          &   $\mu \mathbb{S}^+ = - ( i \mathbb{F}^+ / v_0 + \lozenge )^\ast \circ \mathbb{F}^+$         \\
composite gravitational equations               &   $\mu_g \mathbb{S}^+_g = - \lozenge^\ast \circ \mathbb{F}^+_g$                              \\
composite electromagnetic equations             &   $\mu_e \mathbb{S}^+_e = - \lozenge^\ast \circ \mathbb{F}^+_e$                              \\
composite linear momentum                       &   $\mathbb{P}^+ = \mu \mathbb{S}^+ / \mu_g$                                                  \\
composite angular momentum                      &   $\mathbb{L}^+ = ( \mathbb{R} + k_{rx} \mathbb{X} )^\times \circ \mathbb{P}^+ $             \\
composite octonion torque                       &   $\mathbb{W}^+ = - v_0 ( i \mathbb{F}^+ / v_0 + \lozenge ) \circ \mathbb{L}^+$              \\
composite octonion force                        &   $\mathbb{N}^+ = - ( i \mathbb{F}^+ / v_0 + \lozenge ) \circ \mathbb{W}^+$                  \\
\hline\hline
\end{tabular} \label{tab:3}}
\end{table}

\section{Electromagnetic media}

In the electromagnetic media, the definition of composite electromagnetic equations can be expanded further. It is able to deduce a few corollaries consistent with the Maxwell's electromagnetic theory in the electromagnetic media.

In the second subspace $\mathbb{H}_{em}$ , the composite electromagnetic equations, Eq.(\ref{equ:12}), in the electromagnetic media, can be written as,
\begin{eqnarray}
\mathbb{S}_e + ( k_{fl} / \mu_e ) \mathbb{Z}_e = - \lozenge^\ast \circ ( \mathbb{F}_e + k_{fl} \mathbb{L}_e ) / \mu_e    ~ ,
\label{equ:13}
\end{eqnarray}
in the above, the term, $( \mathbb{F}_e + k_{fl} \mathbb{L}_e ) / \mu_e$ , in the right side can be separated into,
\begin{eqnarray}
( \mathbb{F}_e + k_{fl} \mathbb{L}_e ) / \mu_e
=  && ( \textbf{F}_0 + k_{fl} \textbf{L}_{20} ) / \mu_e + i ( \textbf{E} / v_0 + k_{fl} \textbf{L}_2^i ) / \mu_e
\nonumber
\\
&& ~~~~~ + ( \textbf{B} + k_{fl} \textbf{L}_2 ) / \mu_e
\nonumber
\\
= && i v_0 \textbf{D} + \textbf{H}   ~  ,
\end{eqnarray}
where the paper chooses the gauge condition, $\textbf{F}_0 + k_{fl} \textbf{L}_{20} = 0$ , in the electromagnetic media, for convenience and simplicity. $\textbf{D} = ( \textbf{E} / v_0 + k_{fl} \textbf{L}_2^i ) / ( v_0 \mu_e )$, is the electric induction intensity. $(k_{fl} \textbf{L}_2^i ) / ( v_0 \mu_e )$ is the electric polarization intensity, which contains the electric moment per unit volume of the electric medium. $\textbf{H} = ( \textbf{B} + k_{fl} \textbf{L}_2 ) / \mu_e$ , is the magnetic field intensity. $( - k_{fl} \textbf{L}_2 / \mu_e )$ is the magnetization intensity, which covers the magnetic moment per unit volume of the magnetic medium. The essential component of the term $\textbf{L}_2$ is $( \mu_e \textbf{r} \times q \textbf{V} / \mu_g )$. Compared with the Maxwell's electromagnetic theory in the electromagnetic media, it is found that the coefficient is, $k_{fl} = - \mu_g$ .

Therefore, the composite electromagnetic equations, Eq.(\ref{equ:13}) , can be further written as,
\begin{eqnarray}
\mathbb{S}_e^+ = - \lozenge^\ast \circ ( i v_0 \textbf{D} + \textbf{H} )  ~ ,
\label{equ:15}
\end{eqnarray}
where $\mathbb{S}_e^+ = \mathbb{S}_e - ( \mu_g / \mu_e ) \mathbb{Z}_e$ . $\mathbb{Z}_e = i \textbf{Z}_0 + \textbf{Z}$ . $\textbf{Z}_0 = Z_0 \textbf{I}_0$ , $\textbf{Z} = \Sigma Z_k \textbf{I}_k$ . $\mathbb{S}_e^+ = i \textbf{S}_0^+ + \textbf{S}^+$. $\textbf{S}_0^+ = S_0^+ \textbf{I}_0$ , $\textbf{S}^+ = \Sigma S_k^+ \textbf{I}_k$ . $Z_j$ is real.

Comparing the two sides of the above formula, one can obtain the composite electromagnetic equations in the electromagnetic media as follows,
\begin{eqnarray}
&& \nabla^\ast \cdot \textbf{H} = 0   ~  ,
\label{equ:16}
\\
&& \partial_0 \textbf{H} + v_0 \nabla^\ast \times \textbf{D} = 0    ~  ,
\label{equ:17}
\\
&& v_0 \nabla^\ast \cdot \textbf{D} = - \textbf{S}_0^+    ~  ,
\label{equ:18}
\\
&& \nabla^\ast \times \textbf{H} - v_0 \partial_0 \textbf{D} = - \textbf{S}^+    ~  .
\label{equ:19}
\end{eqnarray}

In the electromagnetic media, the selected gauge condition, $\textbf{F}_0 + k_{fl} \textbf{L}_{20} = 0$, in the paper is distinct from the gauge condition, $\textbf{F}_0 = 0$, in the vacuum. In terms of the electromagnetic media, the gauge condition, $\textbf{F}_0 + k_{fl} \textbf{L}_{20} = 0$, is capable of simplifying the composite electromagnetic equations, Eqs.(\ref{equ:16})-(\ref{equ:19}), for electromagnetic media in the paper. Consequently, the selected Gauss's law for magnetism, in the electromagnetic media, is $\nabla^\ast \cdot \textbf{H} = 0$, rather than $\nabla^\ast \cdot \textbf{B} = 0$ in the vacuum (see Ref.[28]). Obviously, the equation, $\nabla^\ast \cdot \textbf{B} = 0$, is merely one special case of $\nabla^\ast \cdot \textbf{H} = 0$. It is easy to find that the equation, $\nabla^\ast \cdot \textbf{B} = 0$, can be combined with Eqs.(\ref{equ:17})-(\ref{equ:19}) to become the Maxwell's equations for the electromagnetic media, in the classical electromagnetic theory. In other words, the latter can merely be applied to the special case of $\nabla^\ast \cdot \textbf{L}_2 = 0$.

The Gauss's law for magnetism, $\nabla^\ast \cdot \textbf{H} = 0$, in the electromagnetic media, can be further rewritten as follows,
\begin{eqnarray}
\nabla^\ast \cdot \textbf{B} / \mu_e =  \nabla^\ast \cdot ( \mu_g \textbf{L}_2 / \mu_e )    ~  .
\label{equ:20}
\end{eqnarray}

Obviously the above is similar to the equation, $v_0 \nabla^\ast \cdot \textbf{D} = - \textbf{S}_0^+$ . As a result, the term, $\nabla^\ast \cdot \textbf{L}_2 \neq 0$, can be considered as one part of `field source' of the magnetic induction intensity $\textbf{B}$ , in the magnetic media. In other words, the part of `field source' relevant to the magnetization intensity, $( - k_{fl} \textbf{L}_2 / \mu_e )$, will make a contribution on the magnetic induction intensity $\textbf{B}$ , in the magnetic media. Similarly, the part of `field source' relevant to the electric polarization intensity, $( k_{fl} \textbf{L}_2^i ) / ( v_0 \mu_e )$, may make a contribution on the electric field intensity $\textbf{E}$ , in the electric media. These inferences are consistent with that derived from the Maxwell's electromagnetic theory in the electromagnetic media.

The physical properties in the electromagnetic media will be much more colorful and varied than that in the vacuum. Further, it can be considered that some physical properties may exist only in the electromagnetic media. For instance, a few particles and even `magnetic monopole' relevant to the term, $\nabla^\ast \cdot \textbf{L}_2 \neq 0$, may exist merely in some special circumstances, including the electric and magnetic media.

In the second subspace $\mathbb{H}_{em}$ , when the electric polarization intensity, $( k_{fl} \textbf{L}_2^i ) / ( v_0 \mu_e )$, and magnetization intensity, $( - k_{fl} \textbf{L}_2 / \mu_e )$, both can be neglected, the composite electromagnetic equations, Eqs.(\ref{equ:16})-(\ref{equ:19}), in the electromagnetic media, will be degenerated into the Maxwell's equations of electromagnetic fields in the vacuum.

The method of composite electromagnetic equations, in the electromagnetic media, can be extended to the study of composite gravitational equations, in the gravitational media.

\begin{table}[h]
\tbl{Comparison of some Gauss's laws in the material media and vacuum.}
{\begin{tabular}{@{}lll@{}}
\hline\hline
field theory               &   Gauss's laws for magnetism                              &   Gauss's laws for $\textbf{b}$ fields           \\
\hline
vacuum                     &   $\nabla^\ast \cdot \textbf{B} = 0$                      &   $\nabla^\ast \cdot \textbf{b} = 0$             \\
material media             &   $\nabla^\ast \cdot \textbf{H} = 0$                      &   $\nabla^\ast \cdot \textbf{h} = 0$             \\
\hline\hline
\end{tabular} \label{tab:4}}
\end{table}

\section{Gravitational media}

In the octonion spaces, the electromagnetic fields consist of the electric field $\textbf{E}$ and magnetic field $\textbf{B}$. Similarly, the gravitational fields compose the gravitational acceleration field $\textbf{g}$ and gravitational precessional-angular-velocity field $\textbf{b}$ .

In the octonion field theory, the gravitational equations for the gravitational media are similar to that for the vacuum. And the gravitational equations for the gravitational media also possess their own characteristics. In terms of the gravitational media, there are two physical quantities to be respectively similar to the electric polarization intensity and magnetization intensity within the electromagnetic media. These characteristics enable the gravitational equations for the gravitational media to describe a few gravitational properties within the gravitational media.

In the gravitational media, gravitational equations can be further expanded. It is able to deduce some inferences related to the gravitational media, expanding the scope of application of the gravitational theory.

In the quaternion space $\mathbb{H}_g$ , the composite gravitational equations, Eq.(\ref{equ:11}), in the gravitational media, can be written as,
\begin{eqnarray}
\mathbb{S}_g - \mathbb{Z}_g = - \lozenge^\ast \circ ( \mathbb{F}_g - \mu_g \mathbb{L}_g ) / \mu_g    ~ ,
\label{equ:21}
\end{eqnarray}
in the above, the term, $( \mathbb{F}_g - \mu_g \mathbb{L}_g ) / \mu_g$ , in the right side can be separated into,
\begin{eqnarray}
( \mathbb{F}_g - \mu_g \mathbb{L}_g ) / \mu_g
=  && ( f_0 - \mu_g L_{10} ) / \mu_g + i ( \textbf{g} / v_0 - \mu_g \textbf{L}_1^i ) / \mu_g
\nonumber
\\
&&
~~~~~  + ( \textbf{b} - \mu_g \textbf{L}_1 ) / \mu_g
\nonumber
\\
= && i v_0 \textbf{d} + \textbf{h}   ~  ,
\end{eqnarray}
where the paper chooses the gauge condition, $f_0 - \mu_g L_{10} = 0$ , in the gravitational media, for convenience and simplicity. $\textbf{d}$ and $\textbf{h}$ are respectively called as the gravitational-accelerational induction intensity and gravitational-precessional field intensity, temporarily. The gravitational-accelerational induction intensity is, $\textbf{d} = ( \textbf{g} / v_0 - \mu_g \textbf{L}_1^i ) / ( v_0 \mu_g )$ . The gravitational-precessional field intensity is, $\textbf{h} = ( \textbf{b} - \mu_g \textbf{L}_1 ) / \mu_g$ . The term $(- \textbf{L}_1^i / v_0)$ is called as the gravitational-accelerational polarization intensity temporarily, which is similar to the electric polarization intensity in the electric medium. Sometimes the angular momentum, $\textbf{L}_1$, is also called as the gravitational precession-ization intensity within the gravitational media temporarily, which is similar to the magnetization intensity in the magnetic medium.

Consequently, the composite gravitational equations, Eq.(\ref{equ:21}), can be written as,
\begin{eqnarray}
\mathbb{S}_g^+ = - \lozenge^\ast \circ ( i v_0 \textbf{d} + \textbf{h} )    ~ ,
\end{eqnarray}
where $\mathbb{S}_g^+ = \mathbb{S}_g - \mathbb{Z}_g$ . $\mathbb{Z}_g = i z_0 + \textbf{z}$ . $\textbf{z} = \Sigma z_k \textbf{i}_k$ . $\mathbb{S}_g^+ = i s_0^+ + \textbf{s}^+$ . $\textbf{s}^+ = \Sigma s_k^+ \textbf{i}_k$. $z_j$ is real.

Comparing the two sides of the above formula, one can obtain the composite gravitational equations in the gravitational media as follows,
\begin{eqnarray}
&& \nabla^\ast \cdot \textbf{h} = 0   ~  ,
\label{equ:24}
\\
&& \partial_0 \textbf{h} + v_0 \nabla^\ast \times \textbf{d} = 0    ~  ,
\label{equ:25}
\\
&& v_0 \nabla^\ast \cdot \textbf{d} = - s_0^+    ~  ,
\label{equ:26}
\\
&& \nabla^\ast \times \textbf{h} - v_0 \partial_0 \textbf{d} = - \textbf{s}^+    ~  .
\label{equ:27}
\end{eqnarray}

In the gravitational media, the selected gauge condition, $f_0 - \mu_g L_{10} = 0$, in the paper is distinct from the gauge condition, $f_0 = 0$, in the vacuum. In terms of the gravitational media, the gauge condition, $f_0 - \mu_g L_{10} = 0$, is capable of simplifying the composite gravitational equations, Eqs.(\ref{equ:24})-(\ref{equ:27}), for gravitational media in the paper. Consequently, the selected Gauss's law for gravitational precessional-angular-velocity field, in the gravitational media, is $\nabla^\ast \cdot \textbf{h} = 0$, rather than $\nabla^\ast \cdot \textbf{b} = 0$ in the vacuum. And the equation, $\nabla^\ast \cdot \textbf{b} = 0$, is merely one special case of $\nabla^\ast \cdot \textbf{h} = 0$.

The Gauss's law for gravitational precessional-angular-velocity field, $\nabla^\ast \cdot \textbf{h} = 0$, in the gravitational media, can be further rewritten as follows,
\begin{eqnarray}
\nabla^\ast \cdot \textbf{b} / \mu_g =  \nabla^\ast \cdot ( \mu_g \textbf{L}_1 / \mu_g )    ~  .
\label{equ:28}
\end{eqnarray}

Apparently the above is similar to the equation, $v_0 \nabla^\ast \cdot \textbf{d} = - s_0^+$ . Consequently the term, $\nabla^\ast \cdot \textbf{L}_1 \neq 0$, can be considered as one part of `field source' of the gravitational precessional-angular-velocity $\textbf{b}$ , in the gravitational media. In other words, the part of `field source' relevant to the angular momentum, $\textbf{L}_1$ , will make a contribution on the gravitational precessional-angular-velocity $\textbf{b}$ , in the gravitational media. Similarly, the part of `field source' relevant to the gravitational-accelerational polarization intensity, $(- \textbf{L}_1^i / v_0)$, may make a contribution on the gravitational acceleration $\textbf{g}$ , in the gravitational media. These deductions are helpful to deepen the understanding of the physical properties of gravitational media.

The physical properties in the gravitational media may be much more abundance than that in the vacuum. Further, it can be considered that some physical properties may exist only in the gravitational media. For example, the concept and supposition of `several particles' relevant to the term, $\nabla^\ast \cdot \textbf{L}_1 \neq 0$, may exist just in some special circumstances, including the gravitational media.

In the quaternion space $\mathbb{H}_g$ , when the gravitational-accelerational polarization intensity, $(- \textbf{L}_1^i / v_0)$, and angular momentum, $\textbf{L}_1$, both can be neglected, the composite gravitational equations, Eqs.(\ref{equ:24})-(\ref{equ:27}), in the gravitational media, will be degenerated into the gravitational equations, in the vacuum (see Ref.[28]).

\begin{table}[h]
\tbl{Comparison of some components of field strengths in the gravitational and electromagnetic fields. Especially, two words, gravitational-accelerational and gravitational-precessional, correspond to the electric and magnetic respectively.}
{\begin{tabular}{@{}ll@{}}
\hline\hline
electromagnetic fields in the second subspace $\mathbb{H}_{em}$    &  physical quantity                              \\
\hline
electric field intensity                                           &  $\textbf{E}$                                   \\
magnetic induction intensity                                       &  $\textbf{B}$                                   \\
electric induction intensity                                       &  $\textbf{D}$                                   \\
magnetic field intensity                                           &  $\textbf{H}$                                   \\
electric polarization intensity                                    &  $( - \mu_g \textbf{L}_2^i ) / ( v_0 \mu_e )$   \\
magnetization intensity                                            &  $( \mu_g \textbf{L}_2 / \mu_e )$               \\
\hline
gravitational fields in the subspace $\mathbb{H}_g$                &  physical quantity                              \\
\hline
gravitational acceleration                                         &  $\textbf{g}$                                   \\
gravitational precessional-angular-velocity                        &  $\textbf{b}$                                   \\
gravitational-accelerational induction intensity                   &  $\textbf{d}$                                   \\
gravitational-precessional field intensity                         &  $\textbf{h}$                                   \\
gravitational-accelerational polarization intensity                &  $(- \textbf{L}_1^i / v_0)$                     \\
angular momentum                                                   &  $\textbf{L}_1$                                 \\
\hline\hline
\end{tabular} \label{tab:5}}
\end{table}

\section{Experiment proposal}

In the octonion field theory, the composite electromagnetic equations within the electromagnetic media are different from the electromagnetic equations within the vacuum, especially the Gauss's law for magnetism. Similarly, the composite gravitational equations within the gravitational media are dissimilar to the gravitational equations within the vacuum, especially the Gauss's law for gravitational precessional-angular-velocity field (Table \ref{tab:4}). Making use of the implementation of some experiment proposals, it is able to verify the validity of Eqs.(\ref{equ:16}) and (\ref{equ:24}).

\subsection{Magnetization intensity}

Making a comparison between the composite electromagnetic equations, Eqs.(\ref{equ:16})-(\ref{equ:19}), and the Maxwell's equations for the electromagnetic media in the classical electromagnetic theory, it is found that the two field equations are the same, except for the Gauss's law for magnetism. Because the two gauge conditions, selected by the two field equations, are different from each other.

The above analysis result reveals three main points. a) In the electromagnetic equations for the vacuum, the field strength relevant to the electric fields is only the electric field intensity $\textbf{E}$ , while that relevant to the magnetic fields is merely the magnetic induction intensity $\textbf{B}$ . b) In the composite electromagnetic equations, Eqs.(\ref{equ:16})-(\ref{equ:19}), for the electromagnetic media, the field strength interrelated with the electric fields is only the electric induction intensity $\textbf{D}$ , while that interrelated with the magnetic fields is merely the magnetic field intensity $\textbf{H}$ . c) In terms of the classical electromagnetic theory for the electromagnetic media, the electromagnetic strengths in the Maxwell's equations are mixed. That is, the electromagnetic strengths need both the magnetic induction intensity $\textbf{B}$ and magnetic field intensity $\textbf{H}$.

By comparison with the composite electromagnetic equations, Eqs.(\ref{equ:16})-(\ref{equ:19}), the Maxwell's equations are unable to describe the physical properties of magnetic fields completely, in the classical electromagnetic theory for the electromagnetic media. It can be considered that the Maxwell's field equations are transitional, in the electromagnetic media. In other words, the Maxwell's field equations, in the electromagnetic media, are merely the special cases of the composite electromagnetic equations, Eqs.(\ref{equ:16})-(\ref{equ:19}).

According to Eq.(\ref{equ:20}), if the magnetic induction intensity $\textbf{B}$ and magnetization intensity $( \mu_g \textbf{L}_2 / \mu_e )$ both vary periodically at a single frequency, the frequencies of these two physical quantities in the magnetic media must be identical to each other. Making use of this inference, it is able to refer and modify some existing experimental schemes, measuring and comparing the frequencies of magnetic induction intensity $\textbf{B}$ and magnetization intensity $( \mu_g \textbf{L}_2 / \mu_e )$.

These suggested experiments are able to validate the deduction of equal frequencies, confirming the Gauss's law for magnetism, Eq.(\ref{equ:16}) or Eq.(\ref{equ:20}).

\subsection{Angular momentum}

Making a comparison between the composite gravitational equations, Eqs.(\ref{equ:24})-(\ref{equ:27}), and the gravitational equations for the vacuum, it is found that the two field equations are the same, except for the Gauss's law for gravitational precessional-angular-velocity field. Because the two gauge conditions for the two field equations are different from each other (Table \ref{tab:5}).

The above analysis result shows two key points. a) In the gravitational equations for the vacuum, the field strength relevant to the gravitational acceleration fields is only the gravitational acceleration field $\textbf{g}$ , while that relevant to the gravitational precessional-angular-velocity fields is merely the gravitational precessional-angular-velocity field $\textbf{b}$ . b) In the composite gravitational equations, Eqs.(\ref{equ:24})-(\ref{equ:27}), for the gravitational media, the field strength interrelated with the gravitational acceleration fields is only the gravitational-accelerational induction intensity $\textbf{d}$ , while that interrelated with the gravitational precessional-angular-velocity fields is merely the gravitational-precessional field intensity $\textbf{h}$ .

According to Eq.(\ref{equ:28}), if the gravitational precessional-angular-velocity field $\textbf{b}$ and angular momentum $\textbf{L}_1$ both change periodically at a single frequency, the frequencies of these two physical quantities in the gravitational media must be identical to each other. Making use of this inference, it is able to refer and modify some existing experimental schemes, measuring and comparing the frequencies of gravitational precessional-angular-velocity field $\textbf{b}$ and angular momentum $\textbf{L}_1$ . In terms of a single frequency of astrophysical jet, the frequency of its angular momentum is identical to that of its precessional angular velocity.

These suggested experiments can verify the deduction of equal frequencies, validating the Gauss's law for gravitational precessional-angular-velocity field, Eq.(\ref{equ:24}) or Eq.(\ref{equ:28}).

\section{Conclusions and discussions}

In the octonion spaces, the octonion field strength and angular momentum within the vacuum can be combined together to become one new physical quantity, that is, the composite field strength. From the quaternion operator and octonion composite field strength, it is able to define the composite field source. Further, one can expand the definition of octonion composite field source, achieving the composite electromagnetic equations for the electromagnetic media, and the composite gravitational equations for the gravitational media. By means of the composite field strength and field source, we can deeply understand the physical meanings of each term of the octonion angular momentum in the vacuum.

In the electromagnetic media, we can deduce the composite electromagnetic equations related to the electric induction intensity and magnetic field intensity, exploring a few physical properties of electromagnetic media. Especially, in case the magnetic induction intensity and magnetization intensity both vary periodically at a single frequency, the frequencies of these two physical quantities must be identical to each other in the magnetic media. According to this inference, it is able to measure the frequencies of magnetic induction intensity or magnetization intensity. Some inferences, derived from the composite electromagnetic equations for the electromagnetic media, may coincide with that from the Maxwell's equations for the electromagnetic media, in the classical electromagnetic theory.

According to Eq.(16) or Eq.(20), the term, $\nabla^\ast \cdot \textbf{L}_2 \neq 0$ , is one part of `field source' of the magnetic induction intensity $\textbf{B}$ , within the magnetic media. That is, this part of `field source' relevant to the magnetization intensity, $( - k_{fl} \textbf{L}_2 / \mu_e )$, is able to make a contribution on the magnetic induction intensity $\textbf{B}$, within the magnetic media. Obviously, the term, $\nabla^\ast \cdot \textbf{L}_2 \neq 0$ , plays a similar role as some particles and even `magnetic monopole', within the magnetic media. These physical properties may exist merely within the electromagnetic media.

Similarly, in the gravitational media, one can infer the composite gravitational equations related to the gravitational-accelerational induction intensity and gravitational-precessional field intensity, studying some physical properties of gravitational media. Especially, if the gravitational precessional-angular-velocity and angular momentum both change periodically at a single frequency, the frequencies of these two physical quantities must be identical to each other in the gravitational media. According to this deduction, it is able to measure the frequencies of gravitational precessional-angular-velocity field or angular momentum. The composite gravitational equations within the gravitational media can be degenerated into the gravitational equations within the vacuum.

According to Eq.(24) or Eq.(28), the term, $\nabla^\ast \cdot \textbf{L}_1 \neq 0$, is one part of `field source' of the gravitational precessional-angular-velocity $\textbf{b}$ , within the gravitational media. There are several influencing factors to affect the angular momentum, $\textbf{L}_1$ . Under the circumstances, different and multi frequencies of $\textbf{L}_1$ will induce that of $\textbf{b}$ , resulting in different and multi frequencies of astrophysical jets.

From the perspective of the algebra of octonions, the Maxwell's electromagnetic theory is merely one mixture. Some of its equations come from the electromagnetic theory without any electromagnetic medium, while others come from the electromagnetic theory within electromagnetic media.

It is noteworthy that the paper only explores some simple physical properties of the octonion field equations in the electromagnetic and gravitational media, but it clearly deduces some useful inferences. In the electromagnetic media, the magnetic induction intensity and magnetization intensity will meet the requirements of the Gauss's law for magnetism. Similarly, in the gravitational media, the gravitational precessional-angular-velocity field and angular momentum may accord with the demands of the Gauss's law for gravitational precessional-angular-velocity field. In the future studies, it is going to further study the relationship between the magnetic induction intensity and magnetization intensity, within the electromagnetic media. Meanwhile, we shall explore further the interrelation between the gravitational precessional-angular-velocity field and angular momentum, within the gravitational media.

\section*{Acknowledgments}
The author is indebted to the anonymous referees for their valuable comments on the previous manuscripts. This project was supported partially by the National Natural Science Foundation of China under grant number 60677039.


\end{document}